# Magnetic calibration system with interference compensation

Michal Janosek, *Member, IEEE*, Michal Dressler, Vojtech Petrucha, *Member, IEEE* and Andrey Chirtsov

Czech Technical University in Prague, Faculty of Electrical Engineering, 16000 Prague, CZ

The article describes a novel method for calibrating dc-precise magnetometers in the low field range (± 100 µT), which gives acceptable results even in laboratory conditions with significant magnetic interference. By introducing a closely mounted reference magnetometer and a specific calibration procedure, it is possible to compensate for the external magnetic field disturbances caused e.g. by local transportation operated with dc power supplies. The field compensation occurs only shortly after the calibrating coils are energized. In this case, the leakage of coils magnetic flux to the reference sensor due to the cancellation of the time-varying compensating field was negligible. When using 60-cm coils and reference sensor in 2.5-m distance, we were able to calculate magnetometer gains with a standard deviation of 91 ppm. We show that an overall uncertainty of 0.1% can be achieved.

*Index Terms*—calibration, magnetometer, coil system, interference

## I. Introduction

PRECISE calibrations of Earth's field range (± 100 µT) magnetic field sensors (or magnetometers) are challenging not only due to the Earth's field variations, but also by anthropogenous noise, caused e.g. by traffic or nearby uncompensated direct currents. In the literature, three basic calibrating methods are described.

The first method is the so-called "scalar calibration", where the triaxial magnetometer is rotated in (preferably all) spherical directions in a highly homogeneous and stable Earth's magnetic field [1]. For this purpose, it is often required to travel far away from the city to a place with lowest possible magnetic field gradient. The advantage is that only a precise knowledge of the field amplitude is required - this is provided e.g. by an Overhauser scalar magnetometer. Another disadvantage is that the calibration sensor frame is arbitrary. This method is also not much suitable for sensors with high cross-field error [2][3].

The second possibility is using a precisely calibrated coil system and stable, calibrated current source. If the sensor is s, magnetic field gradient can be tolerated. Usually, the "exciting" field is being alternated in every direction and its magnitude is changed to check for linearity and range errors. If the coil system is calibrated, a reference frame can be provided with respect to the device-under-test casing. This technique can be also used for uniaxial sensors and can provide a traceable calibration with uncertainty derivation.

The third calibration method is a mixture of both scalar and vector one: the so-called "thin-shell" method utilizes a fixed sensor head in the triaxial coil system, but the field is artificially rotated and its scalar magnitude is calculated [4]. From a large set of equations all parameters can be established, as in the scalar method. The main benefit is a fixed sensor and thus tolerance to field inhomogeneity.

A usual way of disturbance canceling with coil systems is to place a reference fluxgate magnetometer very far from the coil facility - tens to hundreds of meters, in order not to be influenced by the mutual cross-talk [5][6]. It is also possible to monitor the ambient field changes with a distant Overhauser magnetometer [7] without active cancellation; however any gradient in the disturbance will deteriorate the results.

The disturbances can be also suppressed if the coil system is running in a closed-loop. The closed-loop systems utilize a precise magnetic field sensor, which governs the system precision [8]-[12]. The feedback loop systems in general suffer from possible mutual influence between the device-under-test (DUT) and the feedback sensor (both can generate disturbing magnetic fields), so off-center placement and gradient estimation is necessary.

In this contribution, we focused on a low-cost calibrating system with just 0.6-m triaxial coils, which is usable for the direct vectorial or thin-shell calibration procedure, and allows for calibration of sensors and magnetometers which could otherwise disturb, or could be disturbed, by any intra-coil closed-loop sensor. The setup was running in a laboratory heavily influenced by neighboring dc-traction traffic and other sources of anthropogenous noise. The sensor used for disturbance cancellation was placed just 2.5 meters away from the coil system given by our laboratory constraints.

## II. Coil System Calibration

### A. Available methods

The most accurate calibrations can be provided by a scalar calibration of the coil system [13] [14] which can provide also its non-orthogonalities. Another option is using NMR, mainly with flowing water [15][16] which allows for very small measurement volume of the pickup-coil. A disadvantage of calibration with scalar sensor lies in the required coil system size – its inhomogeneity across the sensor volume is another source of uncertainty for large scalar sensor volume.

Another option is to use a magnetic flux density standard based on a solenoid precisely wound on a quartz-support [17], in this case the achievable accuracy is about 60-ppm

### B. Calibration of the 60-cm coil system

Our coil system (Fig. 1) with overall 60-cm dimensions comprising of one Merritt-coil-quaternion [18] with high homogeneity and two Helmholtz coil. We have done a 3-D FEM simulation (Cedrat FLUX3D) of the Meritt-coils showing that in an area of 12×12 cm² the inhomogeneity is below 50 ppm – see Fig. 2. Due to size constraints, we calibrated the individual coils with a flowing water NMR magnetometer. The uncertainty of 70 ppm (1 σ) was mainly due to field instability - about 30 ppm can be achieved [19].

### III. CALIBRATION PROCEDURE

#### A. The principle

Although we cannot actively compensate for the Earth's magnetic field (~48,000 nT in Europe) because of the coupling to the 2.5-m distant reference sensor, we show that it is possible to compensate only for the magnetic field variations and/or disturbances. The static value of the Earth's magnetic field is then suppressed by performing two or more measurements with different applied fields. We performed a 3D-FEM simulation and have verified by measurements that a 250 nT field in the coils will create a field of 1 nT at the reference sensor. This weak back-coupling results in a slight degradation of the disturbance compensation effectiveness.

#### B. Reference sensor and its alignment

The reference sensor (a 3-axial fluxgate with 200 samples/s digital output and an effective 20-Hz bandwidth after filtering) is placed just 2.5-m away from the coils.

The reference sensor is roughly oriented in the coil system direction before each calibration. It is difficult to align it precisely with the coil system – for this purpose, we use numerical alignment with another, well-calibrated triaxial fluxgate, which is temporarily placed in the coil system and aligned with its axes. A recording of 10-minutes of both magnetometer outputs is enough to calculate the 3×3 transformation matrix **F** between the reference sensor readings **R** and magnetic field vector **C** in the coil system coordinates (Fig.1) - **C=FR**. To obtain matrix **F**, we utilize the magnetic disturbances Δ**C**, which are assumed to be homogeneous on the 2.5-m distance, i.e. we try to solve:

$$\begin{bmatrix} F_{11} & F_{12} & F_{13} \\ F_{21} & F_{22} & F_{23} \\ F_{31} & F_{32} & F_{33} \end{bmatrix} \begin{bmatrix} \Delta R_x \\ \Delta R_y \\ \Delta R_z \end{bmatrix} - \begin{bmatrix} \Delta C_x \\ \Delta C_y \\ \Delta C_z \end{bmatrix} = \min \quad (1)$$

The matrix F was obtained by least-squares inversion of the recording in MATLAB, i.e. **F** = Δ**R**\Δ**C**. The efficiency of the procedure is shown in Fig. 3 – after finding **F**, we first calculated and compared the two aligned sensor readings (left part of the figure –"uncompensated"). After that the current source has been switched on and only the compensating field values were fed to the coil system, effectively suppressing the external field disturbances (right part of the figure – "compensated"). The field variations, which exceeded 500 nT p-p in the vertical (z) axis, were suppressed by a factor of ~10. We can also see that the disturbance in the vertical axis is almost unipolar: averaging would not bring a significant improvement of the calibration quality.

#### C. DUT placement and alignment

The DUT should be generally well aligned with the coil-system axes. For triaxial sensor heads, this is commonly achieved by observing the orthogonal sensor output crossing zero and relying on the sensor orthogonality, since the readings of an aligned sensor exhibit a flat maximum. The residual misalignment, which can be either due to imperfect coil calibrations or due to the sensor non-orthogonalities, causes however an additional error: 1° of misalignment (or sensor pair non-orthogonality) causes about 150 ppm error. However, it is not really necessary to align the sensor precisely, if the coil system is well calibrated (also, coil systems non-orthogonalities are usually below 0.1° [7] [13]). If two orthogonal coils are subsequently energized creating fields $B_x$ and $B_y$, the sensor with a sensitivity $S_x$, misaligned by an angle α, will produce uncalibrated outputs $O_{xx}$ and $O_{yy}$:

$$\Delta O_{xx} = \Delta B_x \cdot \cos\alpha \cdot S_x \quad [\text{V, T, °, V/T}] \quad (2)$$

$$\Delta O_{yy} = \Delta B_y \cdot \sin\alpha \cdot S_x \quad [\text{V, T, °, V/T}] \quad (3)$$

From these two (or more) equations, it is obviously possible to cancel-out the constant misalignment angle and obtain the true, "aligned" sensitivity $S_x$. It is however necessary to take into account any orthogonal component of the magnetic field which could occur in the area where the sensor is placed, from our 3-D simulation we however see that up to 20° rotation is theoretically possible. If the sensor casing is mechanically pre-aligned with one of the coil axes, the angular deviations are obtained for a defined frame, as opposed to thin-shell and scalar calibrations. This applies also for the two remaining (vertical) misalignment angles.

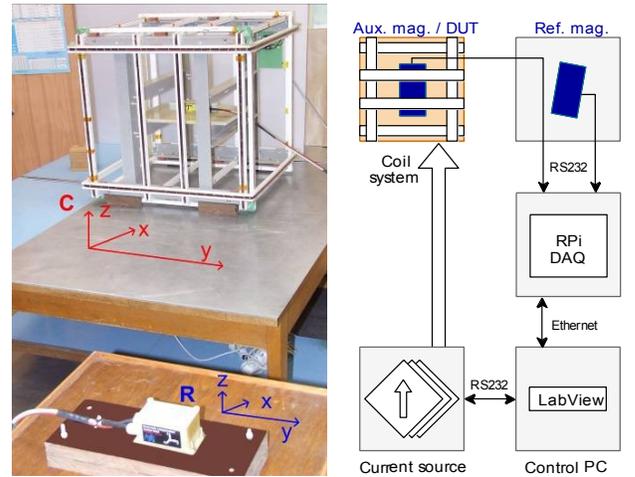

Fig. 1. Left: the coils are wound on fiberglass supports to achieve high temporal and thermal stability. The Merritt-quaternion is located in the "y" (EW) direction. Right: block diagram of the calibration system

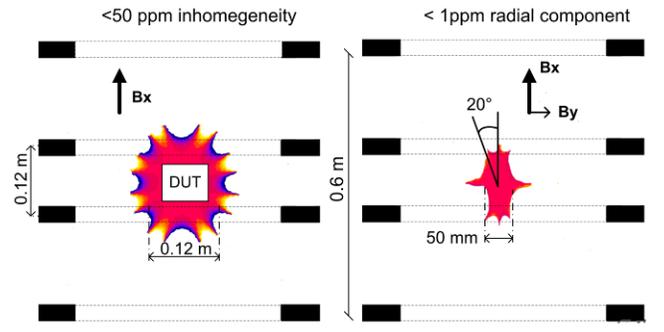

Fig. 2. Left: The red part indicates inhomogeneity below 50 ppm in a 0.12×0.12 m² area where DUT is placed. Right: the red part indicates an area of 0.05×0.12m² where the error due to the radial field component is below 1 ppm, allowing for up to 20° DUT rotation - see (2) and (3).

### D. Current sequencing and field variations compensation

As a very basic calibration method, the currents in the coil systems are sequenced and magnetometer output is recorded, preferably in a bipolar way to suppress the external field variations as much as possible. As we show in Fig. 3, the noise in our laboratory can exceed 500 nT p-p and manifests itself as a unipolar disturbance for most of the time, thus not allowing for efficient averaging. Also DUT settling time is significant: a magnetometer with 1 Hz bandwidth and first-order response settles to 100-ppm in 9 s; achieving 100-ppm stability during this time with a 50,000 nT field equals 5 nT.

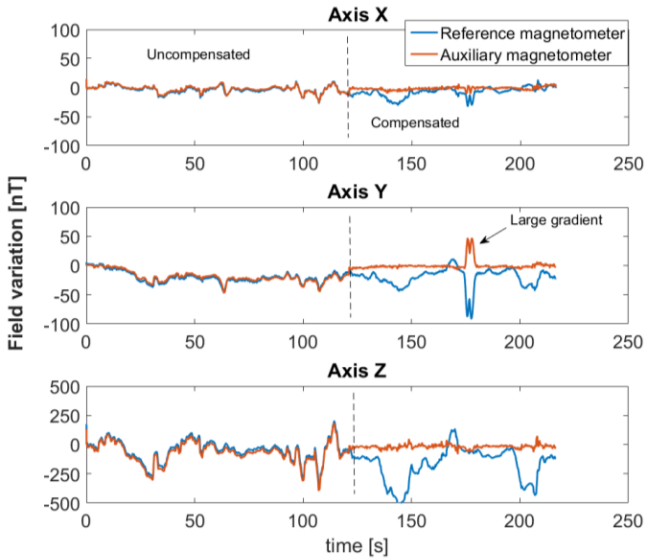

Fig. 3. The field variations sensed by the reference magnetometer (blue – bottom trace) and in the coil system (orange – top trace). The reference magnetometer and the one in the coils have been numerically aligned. The large uncompensated peak was due to steel door opening (large gradient).

Let us describe the calibration procedure for magnetometer gain in the x-axis, assuming the value of α to be zero, the Earth's field x-component $B_{Ex}$ to be static and experiencing time-varying disturbance $B_{Dx}(t)$ during while acquiring $O_x$:

$$O_{x1} = [B_{x1} + B_{Ex} + B_{Dx}(t+t_1)] \cdot S_x \quad (4)$$

$$O_{x2} = [B_{x2} + B_{Ex} + B_{Dx}(t+t_2)] \cdot S_x \quad (5)$$

Obviously, when able to cancel-out the time-varying $B_{Dx}$ term, we can obtain $S_x$ after subtracting Eq. 4 and Eq. 5, which effectively cancels out $B_{Ex}$:

$$O_{x1} - O_{x2} = [B_{x1} - B_{x2}] \cdot S_x \quad (6)$$

In detail, the calibrating sequence for $S_x$ is following:
1. Energize the x-axis coil, creating a calibrating field $B_{x1}$.
2. Wait for settling of the reference magnetometer.
3. In time t=1, record the vector $\mathbf{R}_{(t=1)}$ measured with the reference magnetometer and recalculate it to $\mathbf{C}_{(t=1)}$ in the coil frame using the matrix $\mathbf{F}$ obtained previously: $\mathbf{C}_{(t=1)} = \mathbf{F} \mathbf{R}_{(t=1)}$. The component $C_{x(t=1)}$ is a superposition of calibrating field, Earth's field x-projection and actual value of disturbance in x axis: $C_{x(t=1)} = B_{x1} + B_{Ex} + B_{Dx(t=1)}$.
4. Continually compensate the applied field $B_x$ on the value $C_x(t) - C_{x(t=1)}$, which is in turn only the time-varying disturbance $B_{Dx}(t)$ since $B_x$ and $B_E$ are constant - see (4) and (5). Measure $O_{x1}$ during this time.
5. Repeat for $B_{x2}$, $O_{x2}$ and obtain $S_x$ using equation (6).

The current sequencing is done with a precise 20-bit three-channel current source [20], which is commanded with the calibrating field minus the recalculated disturbance. In Fig. 4 we show one sequence with the short "constant-current" region zoomed-in – the actual disturbance and the residua after suppression are visible.

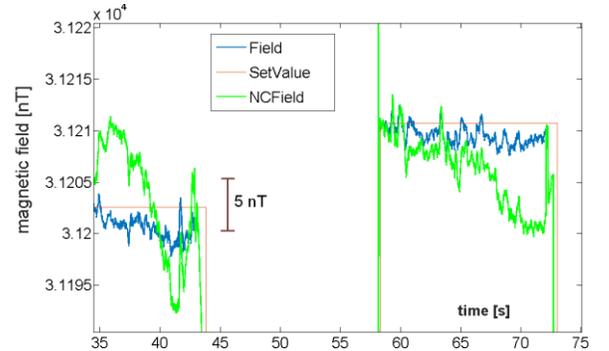

Fig. 4. The uncompensated (green) and compensated (blue) field variations during a calibration sequence with approx. 30,000 nT field step.

### IV. CALIBRATION RESULTS

We have performed several calibration runs where the DUT was a digital triaxial fluxgate magnetometer of our own construction which has been precisely aligned with one of its axis along the y coil axis. The calibrating steps were ± 50 μT and the sensitivity was calculated using (4), (5) and (6).

### A. Standard deviation of the results

By numerically adding the uncompensated magnetic noise, we were able to estimate the improvement of the calibration when using our compensating system – in Fig. 5, the Z-axis differences from mean sensitivity estimation are plotted for both cases - with and without compensation. The improvement is different for different axes – see Table I – in the vertical axis the standard deviation dropped from 1640 down to 311 ppm with compensation switched on. However, for X and Y axes, only ± 50 nT p-p disturbance was observed, so averaging was already effective to suppress the disturbances.

TABLE I
EFFECT OF COMPENSATION ON CALIBRATION

| | Direction | w/o compensation | w/ compensation |
|---|---|---|---|
| Standard deviation - relative [ppm] | NS (x) | 165 | 91 |
| | EW (y) | 134 | 108 |
| | vertical | 1640 | 311 |

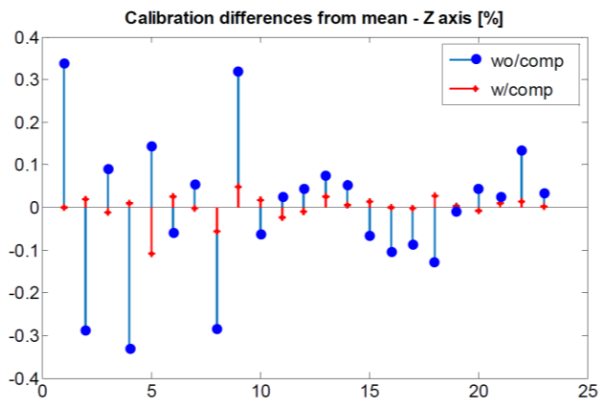

Fig.5. The differences from mean (23 estimations of sensitivity) in the case without (circles) and with (crosses) compensation of field variations

### B. Uncertainty estimation

The above standard deviation of the results is only a small part of overall calibration uncertainty – see Table II. It can be seen, that the largest effect has the angular deviation of the DUT / coil system which for 1 degree misalignment causes an additional 150 ppm. If this angular misalignment is calculated (using equations (1) and (2)) and if the coil system errors are compensated to about 0.1°, this contribution is negligible. The second highest uncertainty source is the coil system transfer constants calibration; we aim to decrease this with a new calibration. The total 1-σ uncertainty for the N-S axis sensitivity is thus either 181 ppm (angular deviations corrected/calculated) or 318 ppm (1 degree misalignment allowance), respectively. Even when allowing for 1 degree error, the expanded 2-σ uncertainty with 95 % probability coverage is below 0.1 % for all three magnetometer axes.

TABLE II
CALIBRATION UNCERTAINTY BUDGET – N-S AXIS

| Uncertainty source | Uncertainty type | 1-σ rel. uncertainty [ ppm ] |
|---|---|---|
| Traceable coil calibration | B | 70 |
| Resistor standard | B | 15 |
| Voltmeter Solartron 7071 | B | 14 |
| Coil inhomogeneity in the central 100x100 mm$^2$ area | B | 50 |
| 1 degree misalignment (when applicable) | B | (0) 150 |
| Current-source noise and tempco during calibration | A | 5 |
| Standard deviation from mean calibration result | A | 91 |
| Current-source nonlinearity | B | 4 |
| TOTAL | A+B | (181) 318 |

### V. CONCLUSION

The main advantage of the presented magnetic calibration system is the low footprint of the compensating system, where the reference magnetometer was placed just 2.5-m away from the coil system, which was allowed by using a novel method of sensor alignment and a calibration sequence. In this manner we are able to cancel disturbing fields which would be inhomogeneous on a larger scale. Since the system does not utilize a feedback-loop, it is possible to calibrate sensors and magnetometers which are producing disturbing magnetic fields or are susceptible to them. In all three axes, we achieved an expanded calibration uncertainty below 0.1 %.

ACKNOWLEDGMENT

This work was supported by the Technology Agency of the Czech Republic, grant TE02000202 "Advanced sensors".